\documentclass[final,1p,times]{elsarticle} 
\usepackage{graphicx} 
\usepackage{amssymb} 
\usepackage{amsthm} 
\usepackage{lineno} 

\usepackage{amsfonts}
\usepackage{amsmath}
\usepackage{amsbsy}
\usepackage{longtable}
\usepackage{bm}
\usepackage{hyperref}
\usepackage{float}

\journal{Nuclear Physics A} 
\begin{document} 

\begin{frontmatter} 


\title{Development of relativistic shock waves in viscous gluon matter}


\author{I.\ Bouras $^{a}$, E.\ Moln\'ar $^{b}$, H.\ Niemi $^{b}$, 
Z.\ Xu $^{a}$, A.\ El $^{a}$, O.\ Fochler $^{a}$, \\ C.\ Greiner $^{a}$
and D.H.\ Rischke $^{a,b}$}

\address[a]{Institut f\"ur Theoretische Physik, 
Johann Wolfgang Goethe-Universit\"at, 
Max-von-Laue-Str.\ 1, D-60438 Frankfurt am Main, Germany}
\address[b]{Frankfurt Institute for Advanced Studies, 
Ruth-Moufang-Str.\ 1, D-60438 Frankfurt am Main, Germany}

\begin{abstract} 
To investigate the formation and the propagation of relativistic
shock waves in viscous gluon matter we solve the relativistic Riemann
problem using a microscopic parton cascade. We demonstrate
the transition from ideal to viscous shock waves by varying
the shear viscosity to entropy density ratio $\eta/s$.
We show that an $\eta/s$ ratio larger than $0.2$ prevents the development
of well-defined shock waves on time scales typical for
ultrarelativistic heavy-ion collisions. These findings are confirmed by
viscous hydrodynamic calculations.
\end{abstract} 

\end{frontmatter} 





\section{Introduction}
Recently, in the context of the jet-quenching phenomena \cite{Adams:2003kv},
exciting jet-associated particle correlations \cite{Wang:2004kfa}
have been observed in heavy-ion collisions at BNL's Relativistic Heavy-Ion
Collider (RHIC). This indicates the formation of shock waves in form
of Mach Cones \cite{Stoecker:2004qu} induced by supersonic partons
moving through the quark-gluon plasma (QGP) and could give a direct access
to the equation of state of the QGP.

Shock waves can be observed if the matter, where the shock waves are 
induced, behaves like a fluid. The large measured elliptic flow 
coefficient $v_2$ \cite{Adler:2003kt} implies that the QGP created at
RHIC could be a nearly perfect fluid with a small viscosity.
Calculations of viscous hydrodynamics \cite{Luzum:2008cw} and microscopic
transport theory \cite{Xu:2007jv} 
have estimated the shear viscosity to the entropy density ratio
$\eta/s$ to be less than 0.4. Still, it is not known, if this upper
limit of the $\eta/s$ ratio is sufficiently small to allow the formation
of relativistic shock waves.

In this work we address the question, whether and when relativistic
shock waves can develop in viscous gluon matter for given $\eta/s$ values.
For this purpose we consider the relativistic Riemann problem. Its
initial conditions are two regions of equilibrated matter with different
constant pressure separated by a membrane which is removed at $t=0$.
The matter evolves in form of a one-dimensional expansion.
We solve the relativistic Riemann problem employing both the parton 
cascade BAMPS (Boltzmann Approach of MultiParton Scatterings) \cite{Xu:2004mz}
and the vSHASTA approach (viscous SHArp and Smooth Transport 
Algorithm) \cite{Molnar:2009tx} of viscous hydrodynamics.
We demonstrate the transition from ideal shock waves to free diffusion 
by varying the $\eta/s$ ratio from zero to infinity. We estimate the 
upper limit of the $\eta/s$ value, for which shocks can still be 
observed experimentally on the time scale typical at RHIC.

\section{BAMPS and vSHASTA}
BAMPS is a microscopic transport model solving the Boltzmann equation
\begin{equation}
p^{\mu} \partial_{\mu} f(x,p) = C(x,p)
\end{equation}
for on-shell gluons with the collision integral $C(x,p)$.
The algorithm for collisions is based on the stochastic interpretation 
of the transition rate \cite{Xu:2004mz}. In this study, we consider
only binary gluon scattering processes with an isotropic
cross section, which is adjusted locally at each time
step to keep a constant $\eta/s$ 
value \cite{Bouras:2009nn,Xu:2007ns,Huovinen:2008te,El:2008yy}.

vSHASTA solves the Israel-Stewart (IS) equations of dissipative
hydrodynamics. In 1+1 dimensions IS equations reduce to \cite{Bouras:2009nn}
\begin{eqnarray}
\label{cons_e}
&&\partial_t T^{00} + \partial_z (v T^{00})\!\! = 
\!\!- \partial_z (v P + v \tilde \pi) , \\
\label{cons_m}
&& \partial_{t} T^{0z} + \partial_z (v T^{0z}) \!\! =
\! \! - \partial_z (P + \tilde \pi) , \\
\label{cons_pi}
&&\gamma \partial_{t} \tilde \pi + \gamma v \partial_z \tilde 
\pi \! \!= \! \!\frac{1}{\tau_{\pi}} \!\left(\pi_{NS} - 
\tilde \pi \right) - \frac{\tilde \pi}{2} \! 
\left(\!\theta + D \ln \frac{\beta_2}{T}\!\right) ,
\end{eqnarray}
where $\theta \equiv \partial_{\mu} u^{\mu}$ and 
$D \equiv u^{\mu} \partial_{\mu}$.
For vanishing shear viscous pressure, $\tilde \pi$, Eqs. (\ref{cons_e})
and (\ref{cons_m}) reduce to the equations of ideal hydrodynamics.

\section{Development of shock waves}
\begin{figure}[th]
\includegraphics[width=13cm]{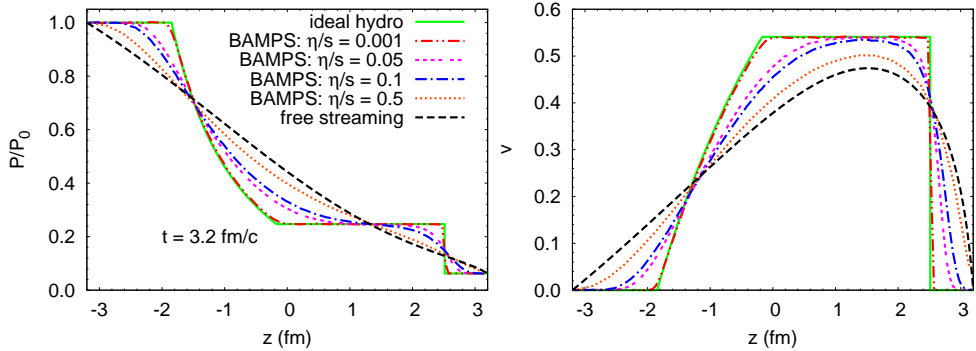}
\caption{(Color online) The solution of the Riemann problem.
At $t=0$, the pressure is $P_0=5.43 \ {\rm GeV fm}^{-3}$ for
$z<0$ and $P_4 =0.33 \ {\rm GeV fm}^{-3}$ for $z>0$.
The left panel shows the pressure and the right panel
the velocity at time $t=3.2$ fm/c.
}
\label{fig01}
\end{figure}

Figure \ref{fig01} shows the solution of the relativistic
Riemann problem for various $\eta/s$ values as computed with
BAMPS. 
The results demonstrate a gradual transition from ideal
hydrodynamic limit ($\eta=0$) to free streaming of particles
($\eta = \infty$).
The ideal solution \cite{Schneider:1993gd} is reproduced with 
a very good precision for $\eta/s=0.001$. We see that a shock wave
is traveling into matter with lower pressure (on the right) with 
a velocity $v_{\rm shock}$ larger than the velocity of sound $c_s$.
The region behind
the shock front is the so-called shock plateau, where matter is
equilibrated and moves collectively with $v_{\rm plat}$,
shown on the right panel of Fig.\ \ref{fig01}. Simultaneously
a rarefaction wave is propagating with $c_s$ into
matter of higher pressure (on the left).

A larger $\eta/s$ value results in a finite transition layer where
the quantities change
smoothly rather than discontinuously as in the case of a
perfect fluid. Furthermore a non-zero viscosity, if large enough,
impedes the formation of a shock plateau and a clear
separation of the shock front from the rarefaction fan.

\begin{figure}[ht]
\includegraphics[width=13cm]{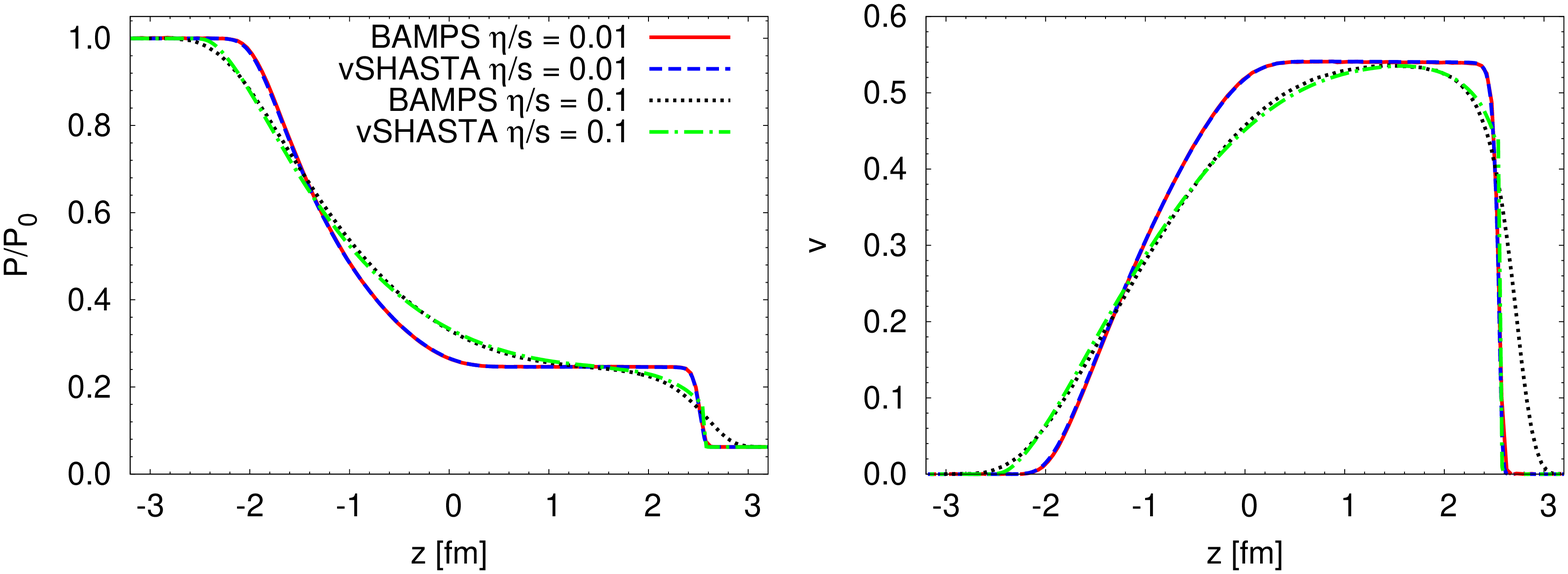}
\caption{(Color online) Same as in Fig. \ \ref{fig01}.
Results are obtained using BAMPS and vSHASTA.}
\label{fig03}
\end{figure}
In Fig.\ \ref{fig03} we compare the results
from BAMPS and vSHASTA for $\eta/s = 0.01$ and $0.1$. We
see a perfect agreement for $\eta/s = 0.01$, whereas for larger
value of $\eta/s = 0.1$ small deviations in the region of the
shock front and rarefaction wave are found. The reason for the difference
is that in these regions the local Knudsen number
$K_{\theta} = \lambda_{\rm mfp} \partial_{\mu} u^{\mu}$ \cite{Betz:2008me}
is large and thus the applicability of IS equations is questionable.
Transport calculations do not suffer from that drawback.

\section{Time scale of formation of shock waves}
\begin{figure}[ht]
\includegraphics[width=13cm]{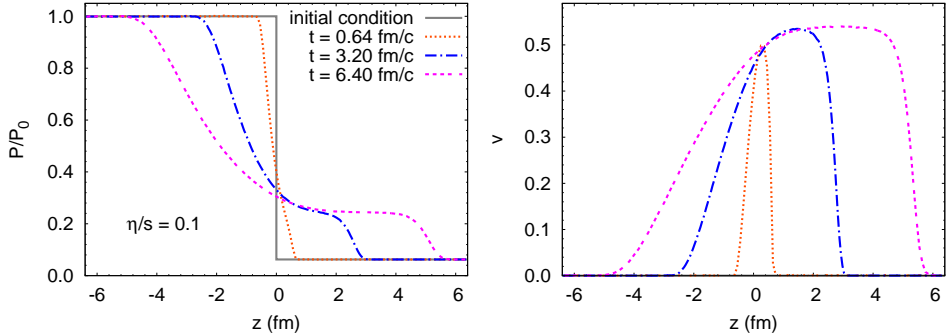}
\caption{(Color online) Same as in Fig. \ \ref{fig01}.
Results are obtained using BAMPS for $\eta/s=0.1$.}
\label{fig02}
\end{figure}
The formation of a shock wave takes a certain amount of time,
as demonstrated in Fig.\ \ref{fig02} for $\eta/s=0.1$.
At early times a shock has not yet developed, the profile looks
like a free streaming of particles. But at later times we observe
the creation of a shock plateau. Formally we define the time of
formation of the shock plateau when
the maximum of the velocity distribution $v(z)$ reaches the value 
$v_{\rm plat}$ of the ideal-fluid solution in Fig.\ \ref{fig01}.
From the right panel of Fig.\ \ref{fig02}, we see that this
happens at $t=3.2$ fm/c. 

In Ref. \cite{Bouras:2009nn} we have found the scaling behavior
of the Riemann problem: the time scale of the formation of shock
waves $t_f$ is proportional to the $\eta/s$ ratio. From Fig. \ref{fig02}
we also infer that, for $\eta/s>0.1$, a shock plateau has not yet 
developed at $t=3.2$ fm/c, whereas for $\eta/s <0.1$, it has already fully
formed. For $\eta/s > 0.2$ we obtain $t_f > 6.4$ fm/c, which is the 
typical lifetime of the QGP
produced in heavy-ion collisions at RHIC. Therefore, an imperfect
quark gluon fluid with $\eta/s > 0.2$ will prevent the observation of
Mach Cone signals at RHIC.

\section{Conclusion}
Using the parton cascade BAMPS
we have solved the relativistic Riemann problem. 
The transition from ideal-fluid behavior to free streaming
is demonstrated. Numerical results from BAMPS agree well
with those obtained from viscous hydrodynamical calculations using vSHASTA.
Considering the scaling behavior we found that the formation
of shock waves in gluon matter with $\eta/s > 0.2$ probably
takes longer than the lifetime of the QGP at RHIC.

\section*{Acknowledgments}

This work was supported by the Helmholtz International Center
for FAIR within the framework of the LOEWE program 
launched by the State of Hesse.
E.\ Moln\'ar acknowledges support by the Alexander von 
Humboldt foundation. The work of H.\ Niemi was supported by
the Extreme Matter Institute (EMMI).



\end{document}